\documentclass[12pt]{article}

\usepackage{latexsym}
\usepackage{amsfonts}
\usepackage{amssymb}
\usepackage{amsmath}
\usepackage{epsf}

\begin{document}

\author{A. S. Sanz \\
 Chemical Physics Theory Group, Department of Chemistry, \\
 University of Toronto, Toronto, Canada M5S 3H6}

\title{A Bohmian approach to quantum fractals}

\date{}

\maketitle

\begin{abstract}
A quantum fractal is a wavefunction with a real and an imaginary part
continuous everywhere, but differentiable nowhere.
This lack of differentiability has been used as an argument to deny
the general validity of Bohmian mechanics (and other trajectory--based
approaches) in providing a complete interpretation of quantum
mechanics.
Here, this assertion is overcome by means of a formal extension of
Bohmian mechanics based on a limiting approach.
Within this novel formulation, the particle dynamics is always
satisfactorily described by a well defined equation of motion.
In particular, in the case of guidance under quantum fractals, the
corresponding trajectories will also be fractal.
\end{abstract}


\section{Introduction}
 \label{sec1}

Quantum mechanics is the most powerful theory developed up to now to
describe the physical world.
However, its standard formulation, based on statistical grounds, does
not provide an intuitive insight of microscopic phenomena as classical
mechanics does for macroscopic ones.
For example, the evolution of a system cannot be followed in the
configuration space by means of well defined, individual trajectories.
On the contrary, the wavefunction associated to such a system extends
to the whole available space, describing the probability for the system
to be located at each (space) point at a certain time.

In order to obtain a more intuitive picture of quantum phenomena,
alternative approaches relying on the concept of {\it trajectory} have
been proposed \cite{Sanz1}.
One of them is Bohmian mechanics \cite{Bohm1,Duerr1,Holland}.
This theory is not merely a reinterpretation of the standard quantum
mechanics despite its equivalence at a predictive level, but a
generalization of classical mechanics that accounts for quantum
phenomena.
Hence, since Bohmian mechanics formally rests on the same conceptual
grounds as classical mechanics, the notion of trajectory (causally
describing the particle evolution) can also be applied in a natural
way to study of microscopic world without contradicting the
statistical postulates of the standard quantum mechanics.

Recently, it has been argued \cite{Hall1} that trajectory--based
theories, like Bohmian mechanics, fail in providing a {\it complete}
interpretation of quantum mechanics.
In particular, these theories could not satisfactorily deal with
wavefunctions displaying fractal features, the so--called {\it quantum
fractals} \cite{Berry,Hall2,Wojcik}.
The existence of this class of wavefunctions in quantum mechanics has
important consequences from a fundamental viewpoint: despite the
coarse--graining restrictions implied by Heisenberg's uncertainty
principle, these wavefunctions constitute the proof that fractal
objects can appear in quantum mechanics as well as in classical
mechanics.
Indeed, W\'ocik {\it et al.}\ \cite{Wojcik} pointed out that quantum
fractals could be experimentally constructed by considering heavy atoms
or ions in macroscopic traps, where the number of energy levels would
be large enough to observe scaling properties at least up to several
orders of magnitude (what is considered a physical fractal).
On the other hand, Amanatidis {\it et al.}\ \cite{Amanatidis} have
observed theoretically this fractal behaviour during the ballistic and
diffusive evolution of wave packets moving in tight--binding lattices,
a study of interest in quantum information theory and quantum
computation.

Here it is shown that the incompatibility between Bohmian mechanics
and the existence of quantum fractals can be easily avoided by
reformulating the particle equation of motion.
This reformulation is based on a limiting approach, where the
wavefunction is expanded in a series of eigenvectors of the
Hamiltonian.
If the wavefunction is not fractal, but {\it regular}, the
trajectories are the same as directly computed by means of
standard Bohmian mechanics.
However, if the wavefunction presents fractal features, its
non--differentiability forbids a direct calculation of the
trajectories, which can be obtained, on the other hand, by means of
the limiting approach proposed here.
Thus, this formal extension of Bohmian mechanics, based on a novel
reformulation of the particle equation of motion, provides a causal
picture for any arbitrary wavefunction, regular or fractal.

The organization of this paper is as follows.
In order to make the paper self--contained, a survey on quantum
fractals is given in section~\ref{sec2}.
The fundamentals of Bohmian mechanics and its generalization to
deal with quantum fractals are presented in section~\ref{sec3}.
The application of the new concepts introduced in this work is
illustrated in detail in section~\ref{sec4} by means of the problem
of a non--relativistic, spin--less particle of mass $m$ in a
one--dimensional box.
This simple, integrable problem can be considered a paradigm of
fractals appearing under conditions not necessarily related to a
chaotic dynamics \cite{Wojcik}.
In section~\ref{sec5} the question of the unbounded energy for quantum
fractals and its interpretation in terms of quantum trajectories is
discussed.
Finally, the main conclusions derived from this work are summarized
in section~\ref{sec6}.


\section{Quantum fractals}
 \label{sec2}

A general method \cite{Wojcik} to construct quantum fractals with an
arbitrary fractal dimension consists in using the quantum analog of the
Weierstrass function \cite{Mandelbrot}
\begin{equation}
 W(x) = \sum_{r=0}^\infty b^r \sin (a^r x) ,
  \qquad a > 1 > b > 0 , \qquad ab \geq 1
 \label{eq1}
\end{equation}
the paradigm of continuous fractal function.
Thus, for example, in the problem of a particle in a one--dimensional
box of length $L$ (with $0 < x < L$), solutions of the Schr\"odinger
equation can be constructed as
\begin{equation}
 \Phi_t (x;R) =
  A_R \sum_{r=0}^R n^{r(s-2)} \sin (p_{n,r} x/\hbar)
  {\rm e}^{-{\rm i} E_{n,r} t/\hbar}
 \label{eq2}
\end{equation}
with $2 > s > 0$ and $n \geq 2$.
Here, $p_{n,r} = n^r \pi\hbar /L$ and $E_{n,r} = p_{n,r}^2/2m$ are,
respectively, the quantized momentum and the eigenvalue associated to
the eigenvector that corresponds to the quantum number $n' = n^r$; and
$A_R$ is the normalization constant.
The wavefunction (\ref{eq2}) is continuous and differentiable
everywhere, however, the one resulting from the limit
\begin{equation}
 \Phi_t(x) = \lim_{R \to \infty} \Phi_t(x;R)
 \label{eq3}
\end{equation}
is a {\it fractal object} \cite{note1} in both space and time.

This method to generate quantum fractals basically consists (given $s$)
in choosing a quantum number, say $n$, and then considering the series
that contains its powers, $n' = n^r$.
There is another alternative (and related) method \cite{Berry} to
obtain quantum fractals based on the presence of discontinuities in
the wavefunction.
In this case, although the initial wavefunction can be relatively
regular, fractal features emerge due to the perturbation that the
discontinuities cause on the wavefunction along its propagation.

An illustrative example of this kind of generating process is a
wavefunction initially uniform along a certain interval,
$\ell = x_2 - x_1 \leq L$, inside the box mentioned above,
\begin{equation}
 \Psi_0 (x) = \left\{
  \begin{array}{cc}
   \frac{1}{\sqrt{\ell}} , & \qquad x_1 < x < x_2 \\
   0 , & \qquad {\rm elsewhere}
  \end{array} \right. .
 \label{eq4}
\end{equation}
The Fourier decomposition of this wavefunction is
\begin{equation}
 \Psi_0 (x) = \frac{2}{\pi\sqrt{\ell}} \sum_{n = 1}^\infty
  \frac{1}{n} \ \!
  \left[ \cos (p_n x_1/\hbar) - \cos (p_n x_2/\hbar) \right]
  \sin (p_n x/\hbar)
 \label{eq5}
\end{equation}
and its time--evolved form is
\begin{equation}
 \Psi_t (x) = \frac{2}{\pi\sqrt{\ell}} \sum_{n = 1}^\infty
  \frac{1}{n} \ \!
  \left[ \cos (p_n x_1/\hbar) - \cos (p_n x_2/\hbar) \right]
  \sin (p_n x/\hbar) \ \! {\rm e}^{-i E_n t/\hbar} .
 \label{eq6}
\end{equation}
As can be noticed, this wavefunction is equivalent to assume
$r = R = 1$ in (\ref{eq2}), and sum over $n$, from 1 to $N$, obtaining
the quantum fractal in the limit $N \to \infty$.
This equivalence is based on the fact that the Fourier decomposition
of $\Psi_0$ gives precisely its expansion in terms of the eigenvectors
of the Hamiltonian in the problem of a particle in a box.
However, this is not general, since the Fourier decomposition and the
expansion of $\Psi_0$ in a basis of eigenvectors of the Hamiltonian are
not equivalent when $V$ is not constant along $x$.

The fractality of wavefunctions like (\ref{eq3}) or (\ref{eq6}) can
be analytically estimated \cite{Berry} by taking advantage of a result
from Fourier analysis.
Given an arbitrary function
\begin{equation}
 f(x) = \sum_{\kappa = 1}^K a_\kappa {\rm e}^{-{\rm i} \kappa x}
 \label{eq7}
\end{equation}
its real and imaginary part are fractals (and also $|f(x)|^2$) with
dimension $D_f = (5 - \beta)/2$ if its power spectrum asymptotically
(i.e., for $K \to \infty$) behaves as
\begin{equation}
 |a_\kappa|^2 \sim |\kappa|^{-\beta}
 \label{eq8}
\end{equation}
with $1 < \beta \leq 3$.
Alternatively, the fractality of $f(x)$ can also be calculated by
measuring the length, $\mathcal{L}$, of its real or imaginary part (or
$|f(x)|^2$) as a function of the number of terms, $K$, considered in
the generating series (\ref{eq7}).
Asymptotically, the relation between $\mathcal{L}$ and $K$ is given by
\begin{equation}
 \mathcal{L}(K) \propto K^{D_f - 1}
 \label{eq9}
\end{equation}
which diverges for $f(x)$ being a fractal object.
Notice that increasing the number of terms that contribute to $f(x)$
is analogous to measuring its length with more precision, since its
structure is gradually better determined.

A remarkable feature that characterizes quantum fractals is that the
expected value of the energy, $\langle \hat{H} \rangle$, of these
wavefunctions is unbounded.
This is related to the fact that the familiar expression of the
Schr\"odinger equation
\begin{equation}
 {\rm i} \hbar \partial_t \Psi_t (x) = \hat{H} \Psi_t (x)
 \label{eq10}
\end{equation}
does not hold in general \cite{Hall1,Wojcik}, as happens when
$\Psi_t (x)$ is a quantum fractal.
In this case, neither the l.h.s.\ of equation (\ref{eq10}) nor its
r.h.s.\ belong to the Hilbert space.
Hence, the equality is not formally correct, and the applicability
of this equation fails.
On the contrary, since each term of the series satisfies this equation,
the identity
\begin{equation}
 \big[ \hat{H} - {\rm i} \hbar \partial_t \big] \Psi_t (x) = 0
 \label{eq11}
\end{equation}
which also represents the Schr\"odinger equation, still remains valid.
When this happens, $\Psi_t (x)$ is called \cite{Wojcik} a solution of
the Schr\"odinger equation in a ``weak'' sense.


\section{Quantum fractal trajectories}
 \label{sec3}

The fundamental equations of Bohmian mechanics are commonly derived by
writing the system wavefunction in polar form
\begin{equation}
 \Psi_t (x) = \rho_t^{1/2} (x) \ \! {\rm e}^{{\rm i} S_t (x) /\hbar}
 \label{eq12}
\end{equation}
with $\rho_t = |\Psi_t|^2$ being the probability density and $S_t$ the
(real--valued) phase, and substituting it into the Schr\"odinger
equation (\ref{eq10}).
This leads to two (real--valued) couple differential equations
\begin{eqnarray}
 & \displaystyle \frac{\partial \rho_t}{\partial t} + \nabla \cdot
   \displaystyle \left(\rho_t \ \frac{\nabla S_t}{m} \right)=0 &
  \label{eq13} \\
 & \displaystyle \frac{\partial S_t}{\partial t}
   + \frac{(\nabla S_t)^2}{2m} + V + Q_t = 0 .
 & \label{eq14}
\end{eqnarray}
Equation (\ref{eq13}) is a continuity equation that ensures the
conservation of the flux of quantum particles.
On the other hand, equation (\ref{eq14}), more interesting from a
dynamical viewpoint, is a quantum Hamilton--Jacobi equation governing
the motion of particles under the action of a total effective potential
$V_t^{\rm eff} = V + Q_t$.
The last term in the l.h.s.\ of this equation is the so--called
{\it quantum potential}
\begin{equation}
 Q_t =
  - \frac{\hbar^2}{2m}\frac{\nabla^2 \rho_t^{1/2}}{\rho_t^{1/2}} .
 \label{eq15}
\end{equation}
This context--dependent, non--local potential determines together with
$V$ the total force acting on the system.

In the classical Hamilton--Jacobi theory, $S_t$ represents the action
of the system at a time $t$, and the trajectories describing the
evolution of the system correspond to the paths perpendicular to the
constant--action surfaces at each time.
Analogously, since the Schr\"odinger equation can be rewritten in terms
of the Hamilton--Jacobi equation (\ref{eq14}), $S_t$ can be interpreted
as a quantum action satisfying similar mathematical requirements as its
classical homologous.
The classical concept of trajectory emerges then in Bohmian mechanics
in a natural way, defining the particle trajectory as
\begin{equation}
 \dot{x}_t = \frac{\nabla S_t}{m} = \frac{\hbar}{m} \
  {\rm Im} \left[ \Psi_t^{-1} \nabla \Psi_t \right] .
 \label{eq16}
\end{equation}
Since in Bohmian mechanics the system consists of a wave and a
particle, it is not necessary to specify the initial momentum for the
particles, as happens in classical mechanics, but only their initial
position, $x_0$, and the initial configuration of the wavefunction,
$\Psi_0$.
The initial momentum field is predetermined by $\Psi_0$ via equation
(\ref{eq16}), and the statistical predictions of the standard quantum
mechanics are reproduced by considering an ensemble of
(non--interacting \cite{note2}) particles distributed according to the
initial probability density, $\rho_0$.

Equation (\ref{eq16}) is well defined provided that the wavefunction
is continuous and differentiable.
However, this is not the case for quantum fractals.
This is the reason why one might infer {\it a priori} that Bohmian
mechanics is an incomplete theory of quantum motion \cite{Hall1}
unable to offer a trajectory picture for this type of wavefunctions.
This apparent incompleteness can be nevertheless ``bridged'' by taking
into account the decomposition of the quantum fractal as a sum of
(differentiable) eigenvectors of the corresponding Hamiltonian, and
then redefining equation (\ref{eq16}) in a convenient way.

Since regular wavefunctions are particular cases of quantum fractals
for which the fractal and topological dimensions coincide, the new,
generalized equation of motion will be applicable to any arbitrary
wavefunction, $\Psi_t$.
Such a wavefunction can be expressed as
\begin{equation}
 \Psi_t(x;N) =
  \sum_{n = 1}^N c_n \xi_n (x) {\rm e}^{-{\rm i}E_n t/\hbar}
 \label{eq17}
\end{equation}
with $N \to \infty$, and where $\xi_n(x)$ is an eigenvector of the
Hamiltonian with eigenvalue $E_n$; in the case where the wavefunction
is constituted by a limited number $M$ of eigenvectors, $c_n = 0$ for
$n > M$.
Accordingly, the quantum trajectories evolving under the guidance of
(\ref{eq17}) are defined as
\begin{equation}
 x_t = \lim_{N \to \infty} x_N(t)
 \label{eq18}
\end{equation}
with $x_N(t)$ being the solution of the equation of motion
\begin{equation}
 \dot{x}_N (t) = \frac{\hbar}{m} \
  {\rm Im} \bigg[ \Psi_t^{-1}(x;N) \
  \frac{\partial \Psi_t(x;N)}{\partial x} \bigg] .
 \label{eq19}
\end{equation}

Observe that this reformulation of Bohmian mechanics is not totally
equivalent to the conventional one.
The calculation of trajectories is not based on $S_t$, which cannot
be trivially decomposed, in general, in a series of analytic,
differentiable functions, as happens with $\Psi_t$.
Thus, the existence of trajectories is directly postulated taking
into account equations (\ref{eq18}) and (\ref{eq19}) rather than
equation (\ref{eq14}).
For regular wavefunctions both formulations coincide due to the
differentiability of $S_t$.
Whereas, when dealing with quantum fractals, the particle equation of
motion is only well defined within this reformulation, and gives rise
to quantum fractal (QF) trajectories.
The fractal dimension of these trajectories can be determined by means
of equation (\ref{eq9}), now $\mathcal{L}$ referring to the
QF--trajectory length.


\section{A numerical example}
 \label{sec4}

The problem of a highly delocalized particle inside a one--dimensional
box (i.e., a particle with the same probability to be found everywhere
inside the box) illustrates fairly well the concepts described in the
previous sections.
The wavefunction representing the state of such a particle is given by
equation (\ref{eq4}), with $x_1 = 0$ and $x_2 = L$.
Taking this into account, equation (\ref{eq6}) becomes
\begin{equation}
 \Psi_t (x) = \frac{4}{\pi\sqrt{L}} \ \! {\rm e}^{-i E_1 t/\hbar}
  \sum_{n \ {\rm odd}}
  \frac{1}{n} \ \! \sin (p_n x/\hbar) \ \! {\rm e}^{-i \omega_{n,1} t}
 \label{eq20}
\end{equation}
where $\omega_{n,1} = (E_n - E_1)/\hbar$.
In the numerical calculations, $L = m = \hbar = 1$ (in arbitrary
units, a.u.).

The probability density, $\rho_t$, associated to the wavefunction
(\ref{eq20}) is a periodic function of time, with period
$T = 2\pi/\omega_{3,1} = mL^2/2\pi\hbar$.
To show that this is the periodicity of $\rho_t$ is relatively easy.
At $t = T$, the arguments of the interference terms contained in
$\rho_t$ are $\omega_{n,1} T = 2\pi (n^2 - 1)/8$, with $n > 1$.
Indeed, since $n$ is always an odd integer, it can be written
as $n(k) = 2(k - 1) + 3$, with $k \geq 1$, and then
$\omega_{n(k),1} T = k (k + 1) \pi$.
This result shows that at $t = T$ any argument is always an integer
multiple of $2\pi$, and therefore the minimum time elapsed between two
consecutive recurrences is precisely $T = 2\pi/\omega_{3,1}$ ($k = 1$).

Despite the periodicity of $\rho_t$, the wavefunction (\ref{eq20})
is not truly periodic due to the common time--dependent phase,
$\varphi_t = - E_1 t/\hbar$, multiplying the sum (for example, the
wavefunction undergoes a delay of $-\pi/4$ after each period).
This delaying phase is a general feature for any wavefunction
expressible as (\ref{eq17}), but has no consequences from a quantum
trajectory viewpoint.
Equation (\ref{eq18}) is invariant under space--independent factors
added to the phase $S_t$, since
\begin{equation}
 \dot{x}_N[S'_t] = \dot{x}_N[S_t]
 \label{eq21}
\end{equation}
when $S'_t = S_t + s(t)$. Here, in particular, $s(t) = \varphi_t$.
The invariance of the quantum motion with respect to such factors is
consistent with the fact that two wavefunctions that differ in a
phase factor represent the same state in standard quantum mechanics.
From now on $S_t$ will refer to the phase of (\ref{eq20}) without
the factor $s(t)$.

\begin{figure}
 \begin{center}
 \epsfxsize=4.5in {\epsfbox{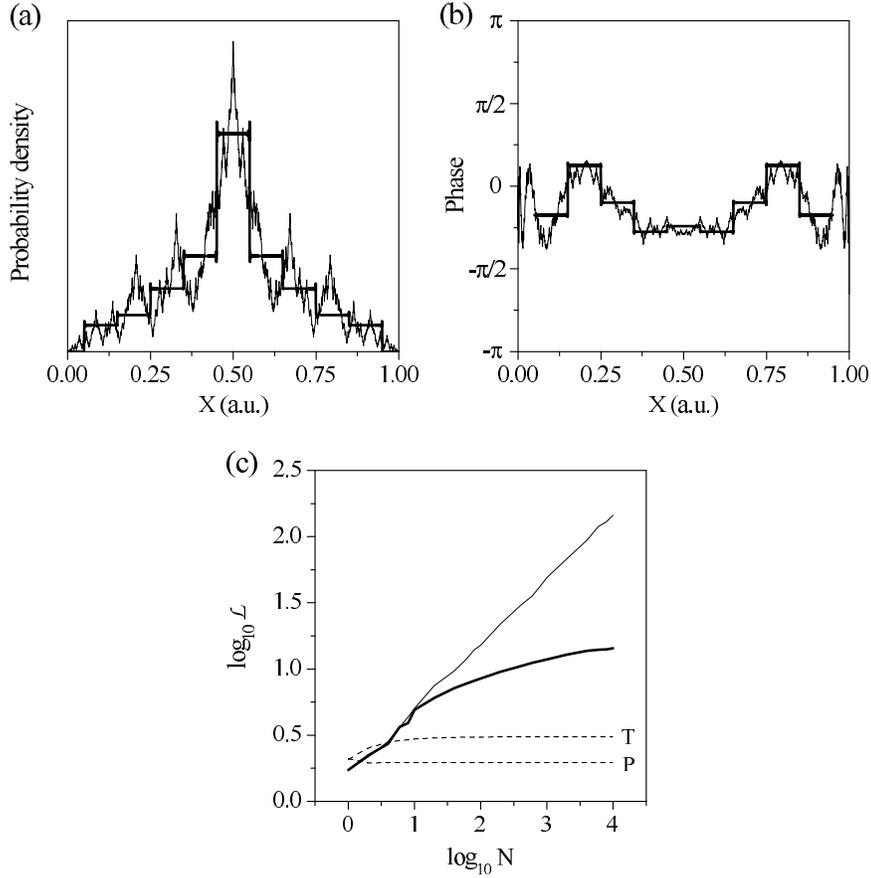}}
 \caption{\label{fig:1} Probability density (a) and phase (b)
  associated to a highly delocalized particle in a box at
  $t = T/\sqrt{2}$ (thin solid line) and $t = 0.7\ \! T$
  (thick solid line).
  (c) Measure of the fractal dimension of the probability densities
  displayed in part (a).
  To compare, measures of the fractal dimension of initial probability
  densities associated to triangular (T) and parabolic (P)
  wavefunctions are also shown.}
 \end{center}
\end{figure}

The profiles along $x$ of $\rho_t$ and $S_t$ are displayed,
respectively, in figures \ref{fig:1}(a) and (b) at two different
times.
These functions display a fractal shape or a revival (characterized by
a step--ladder shape) depending on whether the time is an irrational or
a rational fraction of the period, respectively.
The fractal--revival alternation manifests the Cantor--set structure
\cite{Mandelbrot} of (\ref{eq20}) along time (i.e., its real and
imaginary part display an infinite number of alternating fractal and
revival profiles along time).

The revivals are characterized by the well--known Gibbs phenomenon
related to the Fourier decomposition of discontinuous functions, which
does not affect the quantum motion.
Apart from this, as seen in figure \ref{fig:1}(a), these revivals also
present regions close to the boundaries of the box where $\rho_t$
vanishes at certain times; the most dramatic case happens at $t = T/2$,
when $\rho_t(x) \neq 0$ only in the interval $0.5 < x < 0.75$.
These nodal regions are very important from a dynamical viewpoint.
Since $S_t$ is not well defined in these regions (observe that $S_t$
is not represented for $x \lesssim 0.05$ and $x \gtrsim 0.95$ in figure
\ref{fig:1}(b)), particles avoid them.

The fractal nature of $\Psi_t$ is quantified by applying equation
(\ref{eq9}) to $\rho_t$.
The logarithm of the length ($\mathcal{L}$) of $\rho_t$ as a function
of the logarithm of $N$ is represented in figure \ref{fig:1}(c) for
the two cases considered in part (a).
As clearly seen, $\log_{10} \mathcal{L}$ is proportional to
$\log_{10} N$ in the fractal case, resulting a fractal dimension
$D_f = 1.49$, which is in excellent agreement with that obtained by
Berry \cite{Berry} using equation (\ref{eq8}).
On the other hand, as expected, the length corresponding to the
revival approaches a constant saturation value.
The eventual growth observed in the graph is related to the slow
convergence of $\rho_t$ to a step--ladder structure.
If other regular probability densities with no discontinuities are
considered, the convergence is much faster.
This happens, for example, when one considers that $\Psi_0$ is a
triangle (T) or a parabola (P), both centred at $x_c = 0.5$.
In these cases, also represented in figure \ref{fig:1}(c), the
saturation is reached relatively faster, since only few eigenvectors
are necessary to obtain an excellent convergence.
The slower convergence in the case of $\rho_0^{(T)}$ is due to the
non--differentiability of $\Psi_0^{(T)}$ at $x_c$, which implies a
larger number of eigenvectors in the sum.

\begin{figure}
 \begin{center}
 \epsfxsize=4.5in {\epsfbox{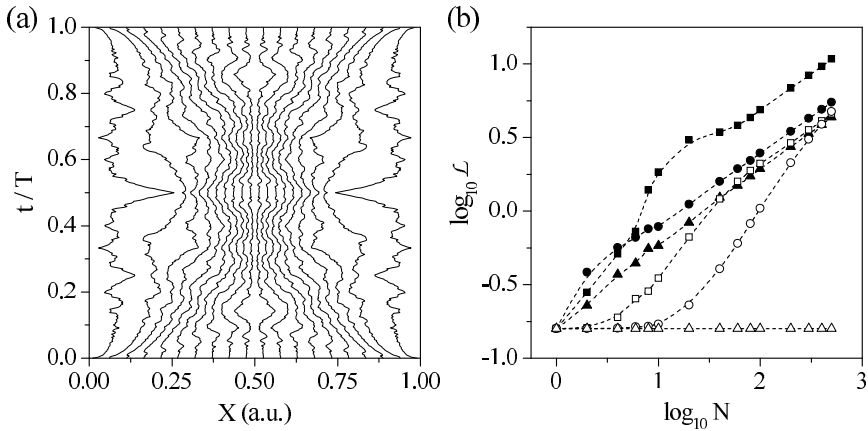}}
 \caption{\label{fig:2} (a) QF--trajectories associated to a highly
  delocalized particle in a box.
  (b) Measure of the fractal dimension of a sample of QF--trajectories
  with initial positions:
  $x_0 = 0.01$ ($\blacksquare$), $x_0 = 0.1$ ($\bullet$),
  $x_0 = 0.4$ ($\blacktriangle$), $x_0 = 0.49$ ($\square$),
  $x_0 = 0.499$ ($\circ$), and $x_0 = 0.5$ ($\triangle$).}
 \end{center}
\end{figure}

The complex space--time structure generated by $\rho_t$ along its
evolution, the so--called {\it fractal quantum carpet} \cite{note3},
can be easily understood by means of the QF--trajectories, which
provide a causal description for such a pattern.
As seen in figure \ref{fig:2}(a), these trajectories manifest the
symmetries displayed by $\Psi_t$, the guiding wave.
Thus, in the case of the reflection symmetry with respect to $x_c$, the
trajectories started at one side of the box (to the left or right of
$x_c$) do not ever cross to the other side.
This effect due to the single--valuedness of $S_t$, which avoids the
trajectories to cross at the same time, can be compared with a
hard--wall scattering problem; an ensemble of particles initially
moving towards the wall will display similar features to those observed
in figure \ref{fig:2}(a) (see, for example, \cite{Sanz2,Sanz3}).
Here, the particles cannot cross the point $x_c$, acting like a
fictitious wall, and therefore they bounce backwards describing
trajectories symmetric with respect to $t = T/2$.
This inversion of the particle momentum is related to the second kind
of symmetry that affects the wavefunction propagation: the change of
sign of $S_t$ during the second half of the period.
Moreover, unlike classical trajectories (and also caused by the
single--valuedness of $S_t$), not all QF--trajectories can reach the
wall, but will move parallel to it.
This is a nice manifestation of the effects caused by the {\it quantum
pressure} \cite{Sanz1} under fractal conditions.

In figure \ref{fig:2}(b), the logarithm of the length of several
QF--trajectories, $\log_{10} \mathcal{L}$, is given as a function of
$\log_{10} N$.
As clearly seen, the converge to proportionality is faster for
those QF--trajectories started at intermediate positions, between
the boundary and the centre of the box.
Notice also that, since the trajectory started at $x_c$ is not a
fractal, its length does not depend on $N$.
Independently of the initial position (and with the exception of the
trajectory started at $x_0 = 0.5$), the fractal dimension of any
trajectory asymptotically approaches the same value, $D_f \simeq 1.50$,
which coincides with that found for $\rho_t$.


\section{Causal considerations about the infiniteness
 of $\langle \hat{H} \rangle$}
 \label{sec5}

As seen in section \ref{sec2}, the expected value of the energy
becomes infinite for quantum fractals.
In the wavefunction (\ref{eq2}), for example, the condition
$2 > s > 0$ gives rise to a divergent series in
$\langle \hat{H} \rangle$ when the limit (\ref{eq3}) is taken into
account.
This property, {\it unavoidable} when dealing with quantum fractals,
is related to their infinite scaling behaviour \cite{Wojcik}, which
is lacking in (everywhere and anytime) regular wavefunctions.
In the case of revivals of wavefunctions with discontinuities, like
(\ref{eq20}), which are regular only at certain times,
$\langle \hat{H} \rangle$ also remains unbounded because an infinite
number of eigenvectors is necessary to recreate the discontinuities.
In this way, the discontinuities can be understood \cite{Berry} as
{\it perturbations} that propagate along the box in time, leading to
the formation of the quantum fractals.

A more physical insight on the infiniteness of
$\langle \hat{H} \rangle$ can be gained by appealing to the
trajectory formulation introduced in section~\ref{sec3}.
In Bohmian mechanics, the particle energy is given by
\begin{equation}
 E_t = \frac{(\nabla S_t)^2}{2m} + V + Q_t .
 \label{eq22}
\end{equation}
Except in the case of particles associated to eigenvectors of the
Hamiltonian, $E_t$ does not conserve in time in general, although the
average energy of an ensemble of particles, initially distributed
according to $\rho_0$,
\begin{equation}
 \bar{E} = \int E_t \rho_t dx = \langle \hat{H} \rangle
 \label{eq23}
\end{equation}
does it, in agreement with the standard quantum mechanics.
The classical analog of two couple particles can help to easily
understand this fact; although the energy of each particle varies
along time due to a continuous transfer between both, the total
energy will remain constant.
Therefore, though Bohmian particles are independent \cite{note2},
the presence of the quantum potential in equation (\ref{eq14}) leads
to sort of non--local coupling or dependence between each particle and
the rest from the ensemble (whose evolution is described by equation
(\ref{eq13})).

In the example used in section \ref{sec4} the potential energy is
$V = 0$ at any time.
Thus, the initial total energy of almost all particles is zero, since
they are distributed according to a constant probability density
($Q_0 = 0$) and the wavefunction is real ($S_0 = 0$).
Only at the boundaries of the box $Q_0 = \infty$ due to the
discontinuity.
This infinite amount of energy is stored up in the particles located
at $x = \epsilon$ and $x = L - \epsilon$ with $\epsilon \to 0^+$,
which constitute {\it energy reservoirs}.
These energy reservoirs are the set of particles located at the
discontinuities of $\Psi_t$ whenever a revival emerges, and not only
at $t = 0$.

The dynamical evolution of the system described by the wavefunction
(\ref{eq20}) can be explained in terms of this initial non--homogeneous
energy distribution among particles as a function of their initial
position.
Although the particle distribution is homogeneous in space, the
particles are not in {\it quantum equilibrium}, but subjected to an
infinite gradient of energy at the boundaries of the box.
This gradient leads to a strong, symmetric energy flow going from the
boundaries towards the centre of the box that makes particles to move
as shown in figure \ref{fig:2}(a).

\begin{figure}
 \begin{center}
 \epsfxsize=3.8in {\epsfbox{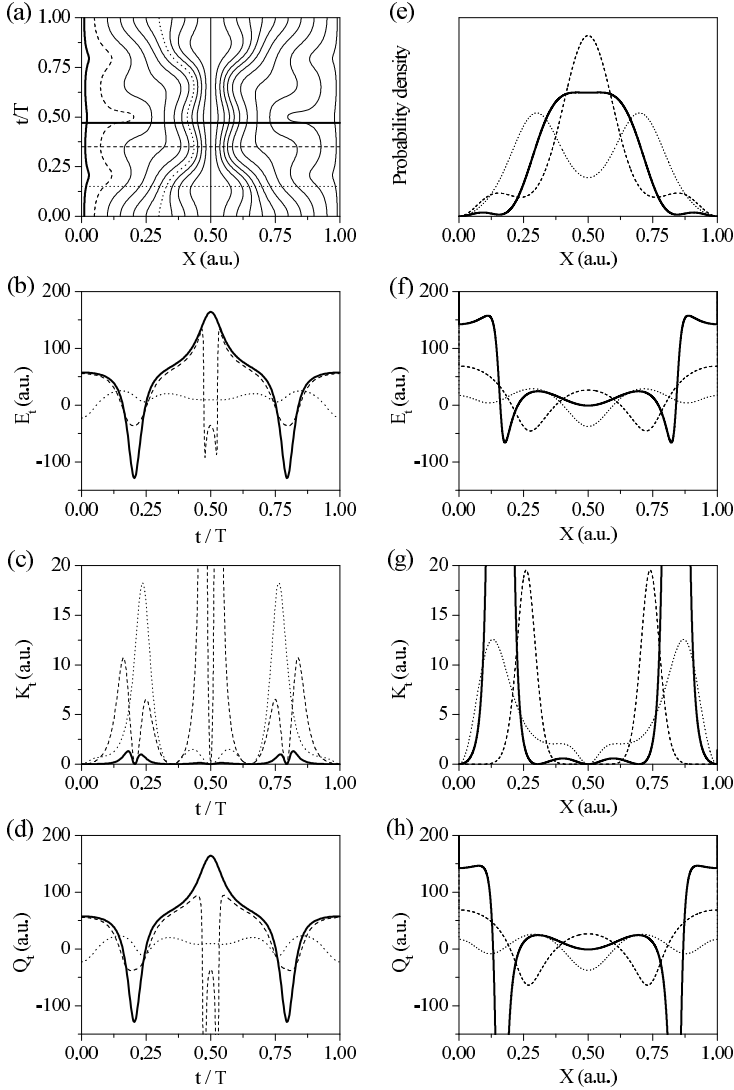}}
 \caption{\label{fig:3} (a) Quantum trajectories associated to the
  wavefunction (\ref{eq20}) with $N = 3$.
  The total and kinetic energy, and the quantum potential are
  represented, respectively, in parts (b), (c) and (d) for three
  different trajectories of panel (a) with initial conditions:
  $x_0^{(1)} = 0.01$ (thick solid line), $x_0^{(2)} = 0.05$ (dashed
  line) and $x_0^{(3)} = 0.3$ (dotted line).
  (e) Probability density at three different times: $t_1 = 0.15$
  (dotted line), $t_2 = 0.35$ (dashed line) and $t_3 = 0.47$
  (thick solid line), in units of $T$.
  These times are indicated in panel (a) by parallel, horizontal
  lines.
  The total and kinetic energy, and the quantum potential at these
  three times are represented, respectively, in parts (f), (g) and
  (h).}
 \end{center}
\end{figure}

The relationship between the time dependence of the energy and the
QF--trajectory dynamics can be better understood, without loss of
generality, by recalling a simpler example consisting in assuming
$N = 3$ in the wavefunction (\ref{eq20}).
An ensemble of trajectories illustrating the dynamics associated with
this case is shown in figure \ref{fig:3}(a); these trajectories can be
regarded as coarse--graining envelopes of the QF--trajectories shown
above.
Although the trajectories have not been distributed according to
$\rho_0$, they provide an insight on how $\rho_t$ evolves, which is
represented in figure \ref{fig:3}(e) at three different times:
$t_1 = 0.15$ (dotted line), $t_2 = 0.35$ (dashed line), and
$t_3 = 0.47$ (thick solid line), in units of $T$.
The total and kinetic energy, and the quantum potential are
represented, respectively, in figures \ref{fig:3}(f)--(h) at these
three times.
On the other hand, these magnitudes are also respectively displayed as
a function of time in figures \ref{fig:3}(b)--(d) for three different
trajectories with initial conditions: $x_0^{(1)} = 0.01$ (thick solid
line), $x_0^{(2)} = 0.05$ (dashed line), and $x_0^{(3)} = 0.3$ (dotted
line).
As seen, the total energy can be, alternatively, positive or negative
along time since it does not conserve.

The quantum potential acting on a particle depends on the structure of
$\rho_t$ at each position of the particle.
Thus, local minima in $\rho_t$ translate into negative wells in $Q_t$
that particles avoid \cite{Sanz4}, moving towards regions with positive
values of $Q_t$, or, at least, presenting local maxima.
The most dramatic case occurs when $\rho_t$ has a node (see $\rho_t$
at $t_3$), manifested as a singularity in $Q_t$.
The intense forces around these regions make particles to move
extremely fast apart from them \cite{note4}, provoking peaks in their
kinetic energies, as seen in figure \ref{fig:3}(g).
This behaviour is not observed, however, for those wells that appear
in the central region.
As commented above, this is because the particle cannot cross the point
$x_c$ in these cases, but keeps moving for a certain time close to it
until the quantum pressure decreases sufficiently, and can move
backwards.

In this way, at $t_1$ very few particles remain close to the borders
of the box (the maxima of $Q_t$ are relatively narrow), and most of
them move towards the maxima located around $x \simeq 0.3$ and
$x \simeq 0.7$, as can be seen in figure \ref{fig:3}(a).
This gives rise to the two important peaks in $\rho_t$; see figure
\ref{fig:3}(e).
At $t_2$, the marginal maxima occupy a wider extension, however they
are relatively high in energy, and therefore most of the particle flow
directs towards the central maxima.
As a consequence, $\rho_t$ displays an important central maximum, and
two secondary, marginal maxima; see figure \ref{fig:3}(e).
Finally, at $t_3$, as seen in figure \ref{fig:3}(e), most of the
particles are collected in the centre of the box, between
$x \simeq 0.25$ and $x \simeq 0.75$, impelled by the strong forces
around $x \simeq 0.16$ and $x \simeq 0.84$, respectively.
Moreover, since $Q_t$ reaches high values at the borders, very few
particles will remain in such regions.

According to this analysis, it is clear that the three kinetic energy
curves shown in figure \ref{fig:3}(g) display peaks on the minima of
$Q_t$, as would happen in a classical situation.
However, the transient trapping observed along $x_c$ has a purely
quantum nature, since there is no physical (classical) potential that
may contribute to it \cite{Sanz2,Sanz3}.
By following the sequence $t_1$--$t_2$--$t_3$, one can see that $Q_t$
progressively increases at the borders and develops deep wells that
confine the particles within the central part of the box.
In other words, the quantum pressure increases from the borders of the
box towards the centre, pushing the particles towards $x_c$, and
obliging them to move along this point for some time (approximately,
half a period).

Taking into account the ideas exposed above, the analysis of a single
particle dynamics becomes much simpler.
For example, as seen in figure \ref{fig:3}(a), the trajectory
$x_t^{(1)}$ is initially slightly pushed away, towards $x_c$, by $Q_t$
until it reaches a turning point, and then moves backwards.
Because of this, two peaks are observed in its kinetic energy, the
second smaller than the first because the particle does not turn back
to the original position.
The turning point, like in classical mechanics, is characterized by a
zero value of $K_t$.
The fact that $Q_t$ reaches its minimum at the turning point can be
understood as an appearance of a non--crossing wall (similar to that at
$x_c$) avoiding the particle to go beyond it.
On the other hand, between $t/T = 0.3$ and $t/T = 0.7$, the particle
remains on top of the plateau seen in figure \ref{fig:3}(h), and it
is almost at rest (there is only a very slight oscillation at about
$t/T = 0.5$).
The same is analysis applicable to the trajectory $x_t^{(2)}$, although
the changes in its velocity are much more relevant, mainly at about
$t/T = 0.5$, when the particle undergoes the strong force due to the
singularity in $Q_t$.
Finally, for the trajectory $x_t^{(3)}$ the double peak is only
observed at half of its evolution unlike the two previous cases.
This is because this particle does not reach any turning point during
the first part of its evolution, but only a sudden force that pushes it
towards $x_c$ in a fast manner.
Once in the trapping region (with the highest quantum pressure) the
particle oscillates, and finally undergoes another sharp forcing that
separates it from the neighbourhood of $x_c$.

In the light of the previous analysis, one can conclude that the
variation in time of the energy can be understood as a regulating
mechanism that adjusts the particle motion in such a way that it
results consistent with the evolution of the wavefunction.
The discussion applied to the trajectories guided by a three--state
wavefunction also remains valid in the case of QF--trajectories.
However, the motion adjustment takes place in a relatively faster
manner, since particles will reach an infinite amount of turning
points along their time evolution.
Therefore, $Q_t$ and $K_t$ will display very deep wells and very
sharp peaks, respectively, and the total (average) energy, given by
equation (\ref{eq23}), will diverge.


\section{Conclusions}
 \label{sec6}

The consistent picture of quantum motion provided by Bohmian mechanics
relies on a translation of the physics contained within the
Schr\"odinger equation into a classical--like theory of motion.
This transformation from one theory to the other is based on the
regularity or differentiability of wavefunctions.
Therefore, it does not hold for quantum fractals, non--regular
solutions of the Schr\"odinger equation.
{\it A priori}, this seems to be a failure of Bohmian mechanics in
providing a complete explanation of quantum phenomena, since quantum
fractals would not have a trajectory--based representation within its
framework \cite{Hall1}.
However, taking into account the fact that Bohmian mechanics is
formally equivalent to the standard quantum mechanics, this
incompleteness results quite ``suspicious''.

By carefully studying the nature of quantum fractals, one can
understand the source of such an incompatibility.
These wavefunctions obey the Schr\"odinger equation in a weak sense
\cite{Wojcik}, i.e., given the wavefunction as a linear superposition
of eigenvectors of the Hamiltonian, the Schr\"odinger equation is
satisfied by each eigenvector, but not by the wavefunction as a whole.
This is because the eigenvectors are always continuous and
differentiable everywhere, unlike quantum fractals, which are
continuous everywhere, but differentiable nowhere.
Taking this into account, a convenient way to express any arbitrary
wavefunction, regular or fractal, is in terms of a superposition of
eigenvectors of the Hamiltonian.
This procedure is particularly important in those circumstances where
the differentiability of the wavefunction is going to be invoked,
like in the formulation of trajectory--based quantum theories like
Bohmian mechanics.

In order to have a truly consistent particle equation of motion,
Bohmian mechanics must be then reformulated in terms of an
eigenvector decomposition of the wavefunction instead of
considering the latter as a whole (as happens in standard Bohmian
mechanics).
The resulting generalized equation of motion, defined by a (convergent)
limiting process, is valid for any arbitrary wavefunction, and
provides the correct Bohmian trajectories.
In the case of quantum fractals, one obtains the desired
trajectory--based picture at the corresponding limit.
Whereas, if the wavefunction is regular, the trajectories determined
by means of this procedure will coincide with those given by the
standard Bohmian equation of motion.
This novel generalization thus proves the formal and physical
completeness of Bohmian mechanics as a trajectory--based approach to
quantum mechanics.

The trajectories associated to quantum fractals are also fractal.
This explains both the formation of fractal quantum carpets and the
unbounded expected value of the energy for quantum fractals.
Although the example of a particle in a box has been used here to
illustrate the peculiarities of quantum fractals, the analysis can be
straightforwardly extended to continuum states \cite{Hall1} or other
trajectory--based approaches to quantum mechanics, like Nelson's theory
of quantum Brownian motion \cite{Nelson}.
Moreover, this kind of analysis can be of practical interest in the
study of properties related to realistic systems, like those suggested
by W\'ocik {\it et al.}\ \cite{Wojcik} and Amanatidis {\it et al.}\
\cite{Amanatidis}, providing moreover a causal insight on their
physics.


\vspace{1cm}
\noindent {\bf \large Acknowledgments}
\vspace{.5cm}

\noindent
The author gratefully acknowledges Prof.\ P.\ Brumer for his support
during the preparation of this work, and Dr.\ M.\ J.\ W.\ Hall for
interesting discussions on the problem posed here.



\begin{thebibliography}{99}
\bibitem{note1}
 Strictly speaking, the wave function (\ref{eq3}) is a semifractal
 \cite{Wojcik} or prefractal \cite{Mandelbrot}, since it is derived
 from a convergent series.
 Prefractals are characterized by having a fractal first derivative.
\bibitem{note4}
 In the case of punctual nodes in high--dimensional problems, a
 vortical dynamics appear \cite{Sanz2}, and the particle can get
 trapped around the quantum vortex temporarily.
\end{thebibliography}
\end{document}